\documentclass{aastex6}
\usepackage{amsmath}
\usepackage{graphicx}
\usepackage{natbib}
\bibpunct{(}{)}{;}{a}{}{,}

\begin {document}
\title{Misaligned disks in the binary protostar IRS~43}
\author{Christian Brinch$^1$, Jes K. J\o rgensen$^2$, Michiel R. Hogerheijde$^3$, Richard P. Nelson$^4$ and Oliver Gressel$^1$}

\affil{$^1$Niels Bohr International Academy, The Niels Bohr Institute,
           University of Copenhagen, Blegdamsvej 17,
           2100 Copenhagen \O, Denmark\\
       $^2$Centre for Star and Planet Formation,
           The Niels Bohr Institute \& Natural History Museum of Denmark,
           University of Copenhagen, \O ster Voldgade 5–-7,
           1350 Copenhagen K, Denmark\\
      $^3$Leiden Observatory,
          Leiden University, PO Box 9513,
          2300 RA Leiden, The Netherlands\\
       $^4$Astronomy Unit,
           Queen Mary University of London, Mile end Road,
           London, E1 4NS, UK
       }

\email{brinch@nbi.ku.dk}

\begin{abstract}
Recent high angular resolution ($\sim 0.2''$) ALMA observations of the 1.1 mm continuum and of HCO$^+$ J=3--2 and HCN J=3--2 gas towards the binary protostar IRS~43 reveal multiple Keplerian disks which are significantly misaligned ($> 60^\circ$), both in inclination and position angle and also with respect to the binary orbital plane. Each stellar component has an associated circumstellar disk while the binary is surrounded by a circumbinary disk. Together with archival VLA measurements of the stellar positions over 25 years, and assuming a circular orbit, we use our continuum measurements to
determine the binary separation, $a=74\pm4$ AU, and its inclination,
$i<30^\circ$. The misalignment in this system suggests that turbulence has likely played a major role in the formation of IRS~43.
\end{abstract}


\section{Introduction}
Planets seem to be a common phenomenon with an estimated occurrence rate of at least one planet per star in the Milky Way~\citep{cassan_one_2012}. Even binary stars are found to have planets orbiting and we currently know about almost one hundred exoplanets orbiting binaries, out of which about ten are circumbinary. A common feature of all these planets around binary stars is that their orbits are well aligned with the orbit of the binary. The largest known deviation from co-planarity is 2.5$^\circ$~\citep{kostov_kepler-413b:_2014}. This alignment suggests that the stars condensed out of a single disk by fragmentation followed by planet formation in the same circumbinary disk through core accretion, a formation scenario which is supported by recent observations \citep{tobin_vla_2013,tobin_vla_2016}. However, there is a strong selection effect because well aligned systems are much easier to detect via transits. Theoretically, \citet{artymowicz_dynamics_1994} and \citet{bate_accretion_1997} showed that for close binaries with separation less than 100 AU, circumstellar disks will form within the central cavity of the circumbinary disk. Such multiple protostellar systems have recently been found, e.g., GG Tau \citep{dutrey_possible_2014} and L1551 IRS~5 \citep{lim_rotationally-driven_2016}, both of which have circumstellar disks as well as a circumbinary disk which are well aligned. \citet{jensen_misaligned_2014} reported misaligned circumstellar disks in the binary star HK Tau, but no circumbinary disk is present in that system. In this paper we report observations of a binary protostar, IRS~43, which also has three separate disks, two circumstellar and one circumbinary, but these are all substantially misaligned ($\gg 5^\circ$). IRS~43 (IRAS 16244-2434) is a binary protostellar system in the Rho Ophiuchus star forming region~\citep[d=120 pc;][]{knude_interstellar_1998}. It is described in the literature as a Class I object due to the shape of the SED~\citep{wilking_iras_1989} and hence is assumed to be of order 100,000 years old. IRS~43 was discovered to be a binary  by \citet{girart_subarcsecond_2000} based on 3.6 cm continuum observations from the VLA taken in 1989. Following these authors we will henceforth refer to the two stellar components as VLA 1 and VLA 2, with VLA 1 being the northernmost source and also the stronger of the two at millimeter and centimeter wavelengths. The binary was later confirmed by observations in the mid--infrared where, curiously enough, VLA 2 is seen to be the stronger source \citep{haisch_infrared_2002,duchene_multiple_2007}.

\section{Observations and data reduction}
IRS~43 was observed by the Atacama Large Millimeter/submillimeter Array (ALMA) on 2015 August 28, spending approximately 1 hour on the source. 36 antennas were in use on that night providing baselines from 15 meters to 1465 meters. Two phase calibrators, J1625--2527 and J1627--2426, were observed approximately every 7 minutes. The precipitable water vapor level was around 2 mm at the time of observing. We used band 6 in our observations and receivers were tuned to 267 GHz (1.1 mm). Of the four spectral windows, one was set up to measure the continuum at 252 GHz using the full 2 GHz bandwidth. The remaining three spectral windows were used to measure the lines HCO$^+$ J=3—-2 at 267.6 GHz, HCN J=3—-2 at 265.9 GHz, and HC$^{18}$O$^+$ J=3—-2 at 255.5 GHz. The latter was not detected. HCO$^+$ was measured at a spectral resolution of 61 kHz corresponding to 68 ms$^{-1}$ while HCN was observed at a slightly lower spectral resolution, 122 kHz corresponding to 138 ms$^{-1}$. Careful recalibration of the data did not improve the quality significantly compared to the delivered version calibrated by ALMA staff, so we settled on using the original version as provided by ALMA. We did experiment with self-calibration of the continuum, but the improvement in signal--to--noise was negligible and did not reveal additional structure, so in the end we decided not to self-calibrate.

In this paper we also make use of archival data from the Very Large Array (VLA) as well as data published by \citet{girart_subarcsecond_2000} and \citet{curiel_very_2003}. Two additional datasets on IRS~43 were found in the NRAO VLA archive, project numbers AC0788 and AF0443, taken in 2005 and 2007 respectively. These two datasets were phase calibrated using CASA and astrometric positions of the protostellar components measured and added to the list of VLA astrometry. In addition we utilize continuum data from ALMA observations of IRS~43 0.4$'' \times$0.3$''$ angular resolution from a larger survey of Class I protostars in Ophiuchus (2013.1.00955.S, PI: Jes J\o rgensen; E.~Artur et al., in prep.)

\section{Results}

\subsection{Dust}
\begin{figure*}[!h]
\begin{center}
\includegraphics[width=8.5cm]{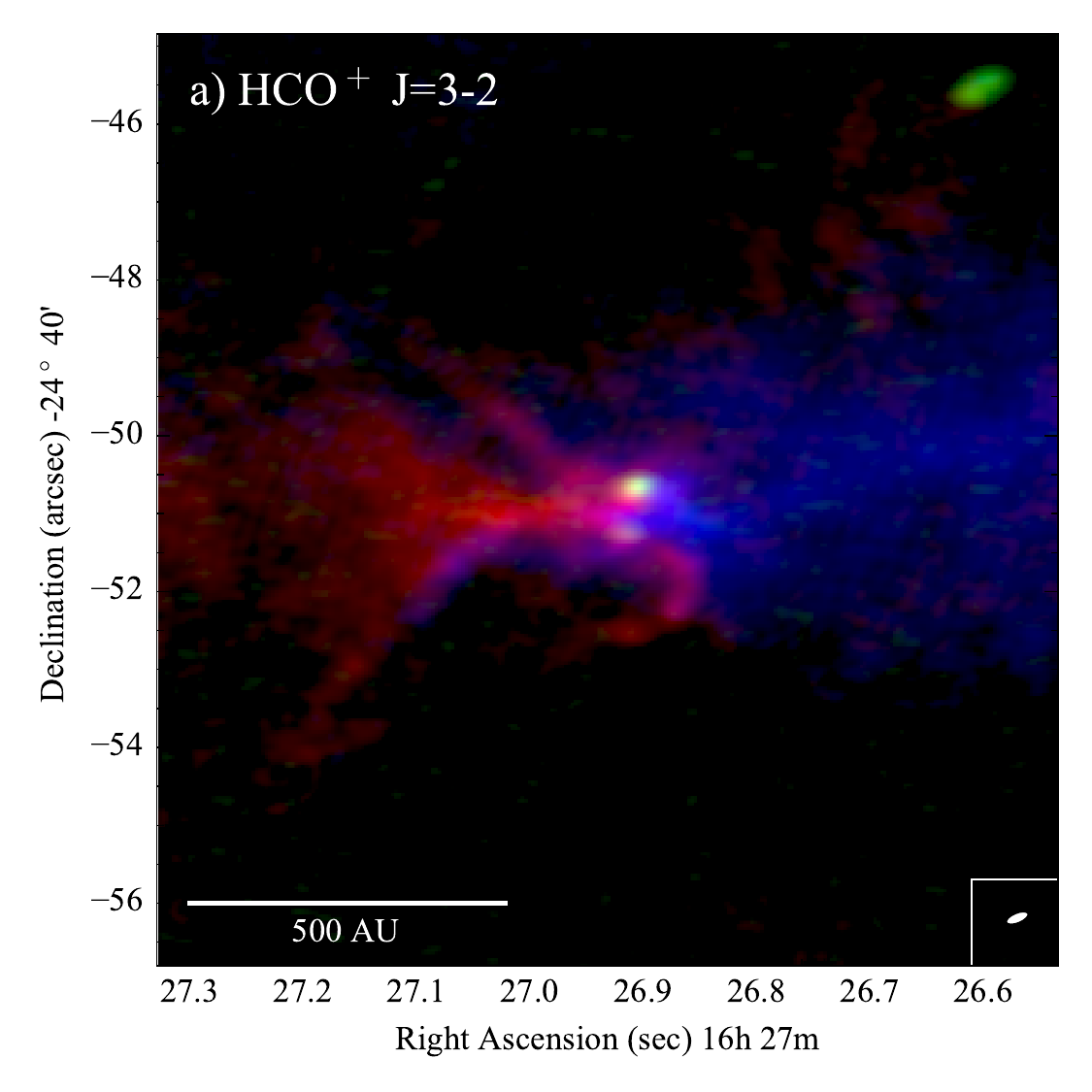}
\includegraphics[width=8.5cm]{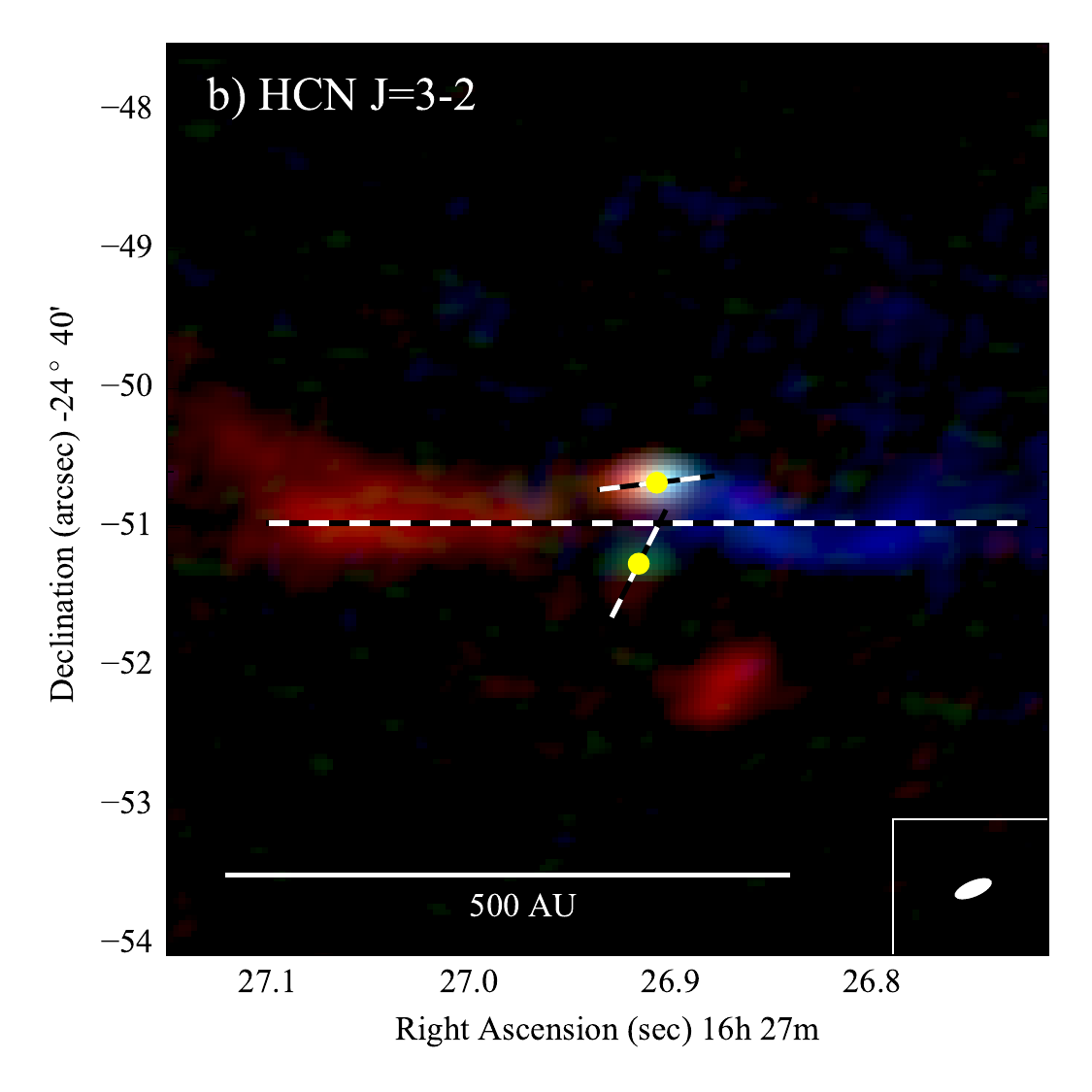}
\end{center}
\caption{Three color images of the continuum and line emission from IRS~43. In both panels, green shows the 1.1 mm continuum, red is the integrated red-shifted line emission and blue is the blue-shifted line emission. Panel a) shows the HCO$^+$ J=3--2 emission and panel b) shows the HCN J=3--2 line. The intensity of all three colors has been normalized by the peak intensity. The white dashed lines in the right panel show the major axes of the three disk components. The positions of the two protostars are marked by yellow circles. }\label{rgb}
\end{figure*}

The continuum emission as observed by ALMA shows two distinct peaks, both of
which are marginally resolved. The continuum is very compact with no trace of extended emission on scales larger than 0.3$''$. The continuum is shown in green color in Fig.~\ref{rgb}, restored with a 0.23$''\times$0.15$''$ beam. A third and very strong continuum source can also be seen in Fig.~\ref{rgb} some 800 AU to the north-east of VLA 1 and VLA 2. This source is known as GY263 and may or may not be gravitationally bound to IRS~43.

The continuum emission originates from cold dust surrounding the the two
protostars. If we assume a constant dust temperature of 15 K and a dust opacity
$\kappa$ at 1.1 mm of 2.7 cm$^2$g$^{-1}$, we can estimate the dust mass around the stars using,
\begin{eqnarray}
M_{dust}= \frac{F_{\lambda} d^2}{\kappa_{\lambda} B_{\lambda}(T_{dust})}
\end{eqnarray}
where $d$ is the distance to the source and $B_\lambda(T)$ is the Planck
function for temperature $T$. The continuum flux of VLA 1 and VLA 2 is 18.5 mJy
and 1.9 mJy respectively, which gives total gas and millimeter dust masses of 770 M$_\oplus$ and 20 M$_\oplus$ after multiplying $M_{dust}$ with the canonical gas-to-dust ratio of 100.

The fact that the continuum is only marginally resolved allows us to place
upper limits to the size of the dust distributions. Given a beam size of
0.3$''$, the dust extends no more than 20 AU from the stars. The dust around
VLA 1 may extend slightly further than that around VLA 2, but not more that
10\%--20\%.

\subsection{Gas}
\begin{figure*}
  \begin{center}
\includegraphics[width=8.0cm]{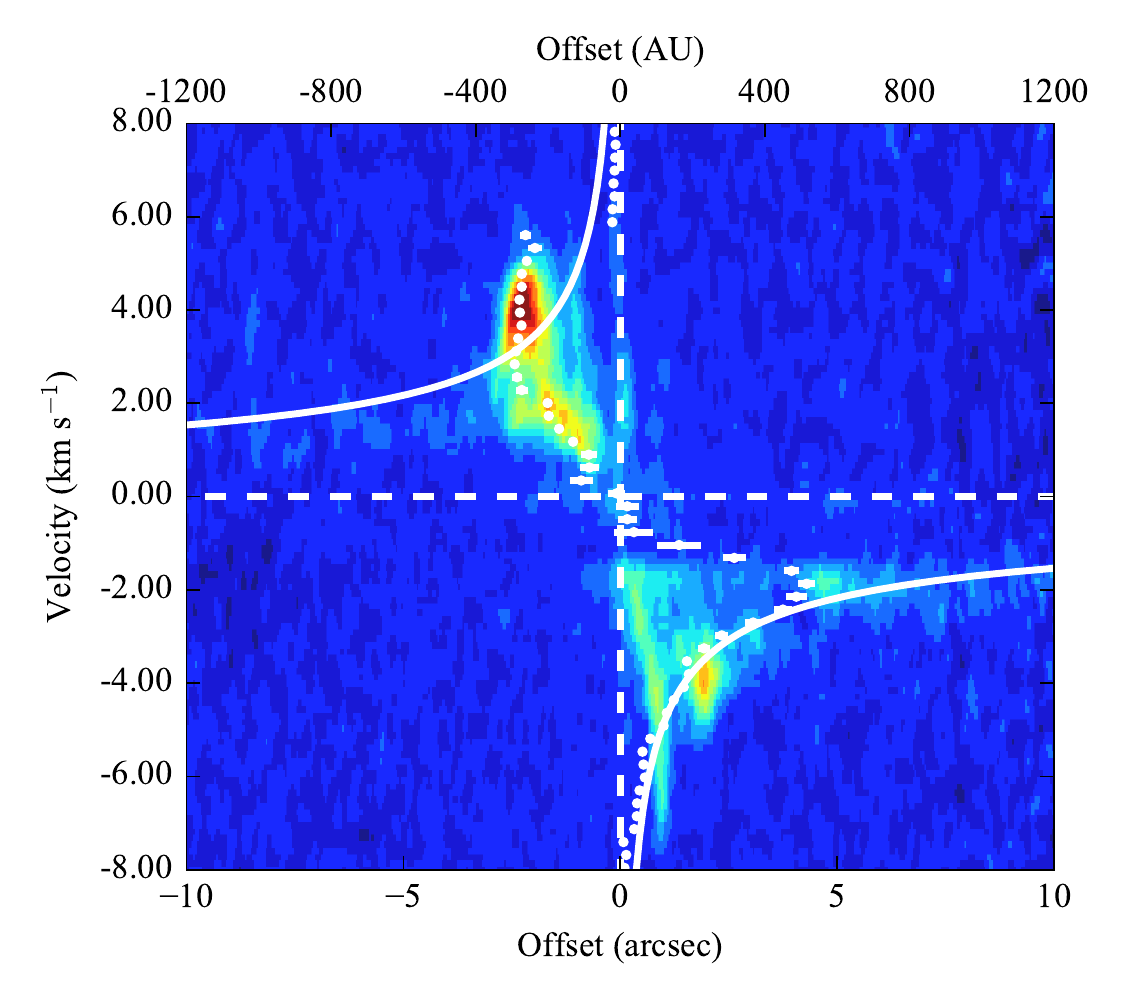}
\includegraphics[width=8.0cm]{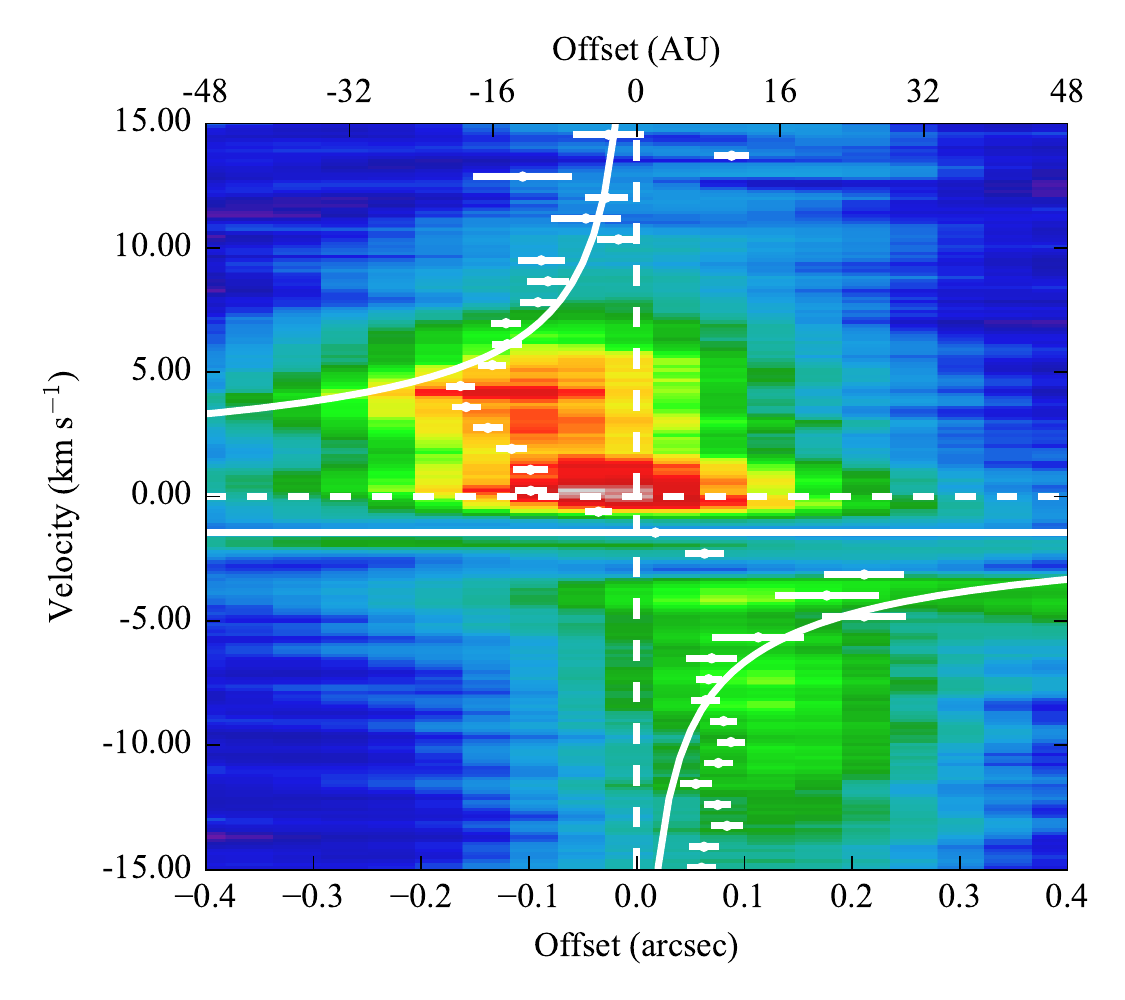}
\end{center}
\caption{PV-diagrams of the HCN J=3--2 emission. The left panel shows the pv-diagram along a cut through the center of mass along the circumbinary disk mid-plane and the right panel shows the pv-diagram along the major axis of the VLA 1 disk exactly through VLA 1 itself. The white markers show the peak position and error bar of a Gaussian fit to the emission in each channel. The full white lines show a Keplerian fit which is discussed in Sect.~\ref{disc}. }\label{pv_hcn}
\end{figure*}

The HCO$^+$ J=3--2 and HCN J=3--2 emission is shown in Fig.~\ref{rgb} along with the continuum emission. Both lines are strong and clear. In these two figures, the integrated red and the blue shifted emission is colored in red and blue. HCO$^+$ is seen to be a lot more extended than HCN, particularly in the direction along the declination, but both tracers show a clear circumbinary red/blue asymmetry. HCO$^+$ shows a X-shaped and slightly curved structure in both red and blue, which is not seen in HCN at all. This structure emits at large velocity offset only ($>$7\ kms$^{-1}$) and is likely associated with an outflow or disk wind. The three disks are easily visible in HCN. In Fig.~\ref{rgb} their axes are marked by white dashed lines. We can confirm that these structures are in fact Keplerian disks, by looking at the velocity field as a function of position offsets. Figure~\ref{pv_hcn} shows the PV-diagrams along the axes shown in Fig.~\ref{rgb} for the circumbinary disk and the circumstellar disk around VLA 1. The signal is not strong enough to prove that the velocity gradient across VLA 2 is Keplerian, but \citet{herczeg_disks_2011} found a disk in CO associated with VLA 2. We therefore assume that the velocity gradient seen in HCN across VLA 2 is due to this disk and we adopt a position angle which lies along this gradient. In both PV-diagrams, a linear segment of emission at low velocities connects the red and the blue side across the center. This is expected for a Keplerian disk of finite radius. The origin of this emission is well explained by~\citet{lindberg_alma_2014}. The blue-shifted side of the circumbinary disk in HCN has an apparent warp.

\subsection{Astrometry}
In 2003, \citeauthor{curiel_very_2003} published a study in which they had followed the proper motion of IRS 43 using the VLA over a 12 year period, from 1990 to 2002. They produced a lower limit to the total mass (1.7 M$_\odot$) and an upper limit on the period (360 yr) as well as an estimate of the binary separation and absolute proper motion of the system.

\begin{figure*}
\includegraphics[width=18cm]{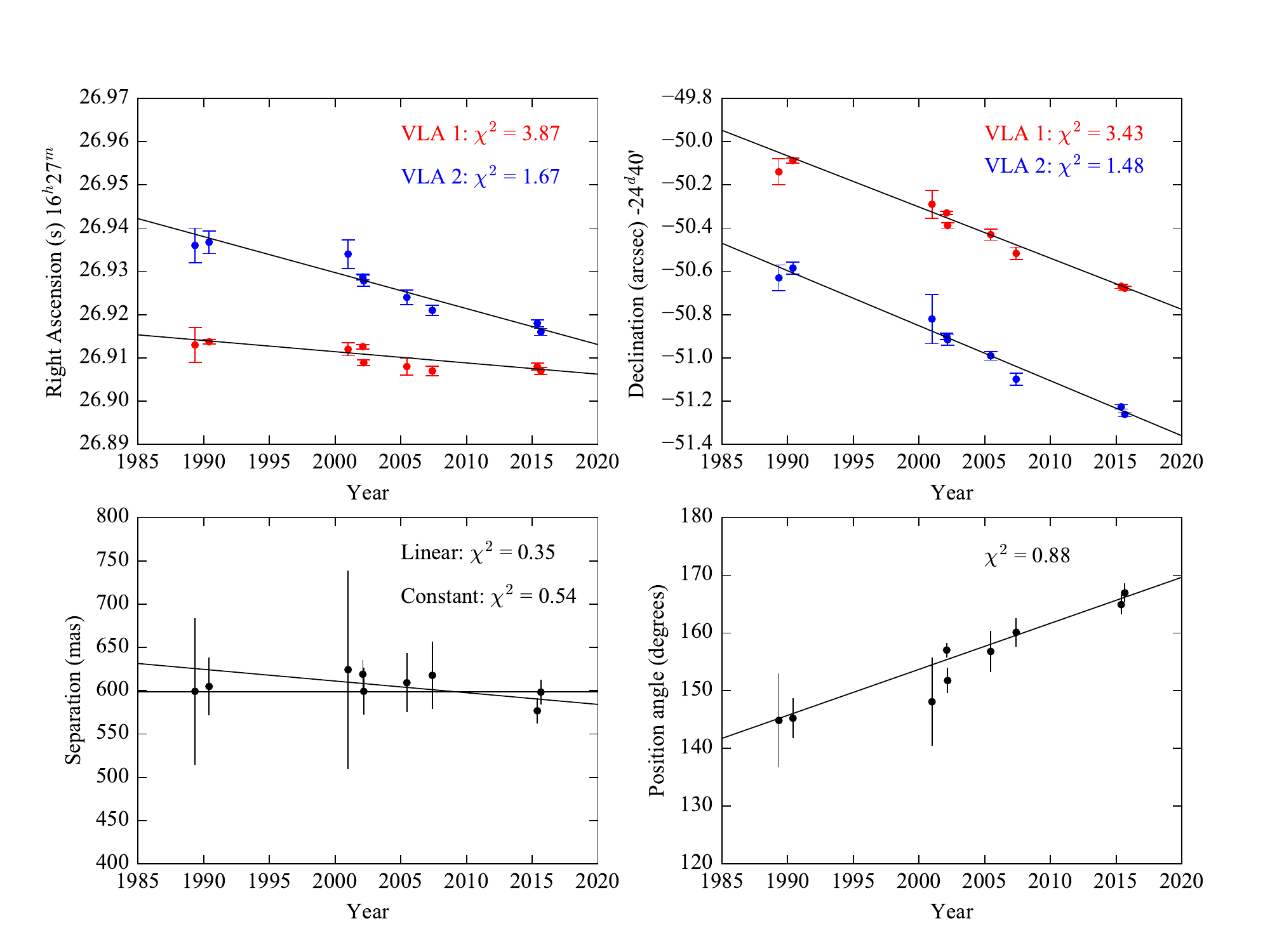}
\caption{The measured positions of VLA 1 and VLA 2 over the span of 25 years (top panels)
         as well as the derived quantities separation and position angle (bottom panels),
         also as function of time. The figure shows the least square fits to the data and
         reduced $\chi^2$ values are given for each fit.}\label{astrometry1}
\end{figure*}

With our recent ALMA data and additional archival VLA data, we can add 13 years
to the fraction of measured orbital period, which allow us to produce a more
accurate determination of the orbital parameters including reliable error bars.
The positions and the derived quantities separation and position angle are shown in Fig.~\ref{astrometry1}. Least square fits to the data are shown as well together with the reduced $\chi^2$ values of each fit. These values show that the fits are very robust. In the case of the binary separation, we have fitted both a linear function, allowing for a change of separation in time, as well as a constant, which would assume a circular orbit in the plane of the sky, i.e., constant separation. Based on the reduced $\chi^2$-values of the two fits, we cannot determine whether the separation of the protostars is constant in time or not, but within the error bars the data is certainly consistent with a circular orbit lying in, or very close to, the plane of the sky. If we allow for a non-zero inclination, we find that the data is inconsistent with an inclination above 30 degrees from the plane of the sky. The least square fit parameters are given in Table~\ref{leastsquarefit} where we give the slope of the best fitting line to the position angle data as the angular velocity.
\begin{deluxetable*}{cccc}
\tablecaption{Best fit parameters\label{leastsquarefit}}
\tablenum{1}
\tablehead{
\colhead{} & \multicolumn{2}{c}{Least squares} & \colhead{\texttt{PIKAIA}} \\
\colhead{Parameter} & \colhead{VLA 1} & \colhead{VLA2} & \colhead{CoM}
}
\startdata
$\mu_{R. A.}$ (mas yr$^{-1}$) &    -3.9 (0.9)   & -12.5 (1.0) & -7.6 (0.5)\\
$\mu_{Dec.}$  (mas yr$^{-1}$) &   -23.6 (0.9)   & -25.4 (0.9) & -25.3 (0.4)\\
Separation    (mas)           &   \multicolumn{2}{c}{599 (6)}  & 615 (37)   \\
Separation    (AU)            &   \multicolumn{2}{c}{72 (1) AU$^a$} & 74 (4) AU$^a$ \\
Period (Years)  &   \multicolumn{2}{c}{451.2 (46.9)} & 444.2 (33.4) \\
Stellar mass ratio & \multicolumn{2}{c}{--} & 0.500 (0.018)\\
Total mass (M$_\odot$)        &   \multicolumn{2}{c}{1.80 (0.42)} & 2.01  (0.47)
\enddata
\tablecomments{$^a$assuming a distance of 120 pc}
\end{deluxetable*}

In order to determine the orbit of the binary, we fit, for each of the two stars
a function of the form
\begin{eqnarray}
\begin{split}
\mathbf{r}(t) = & \mathbf{r}_{CoM;0} + \mathbf{v}_{CoM} t \\
                & + d \ f [\cos(\omega t + \phi_0)\mathbf{\hat{x}}
                         + \sin(\omega t + \phi_0)\mathbf{\hat{y}}]
\end{split}
\end{eqnarray}
where $\mathbf{r}_{CoM;0}$ is the position of the center of mass at $t_0$,
$\mathbf{v}_{CoM}$ is the velocity of the center of mass in the plane of the
sky, $d$ is the projected separation of the stars, $f$ is the mass ratio of the
stars, $\omega$ is the orbital frequency, and $\phi_0$ is the phase (i.e.,
position angle) at $t_0$. A total of 6 free parameters as the position of the
center of mass is fixed by the positions of the stars together with the
parameter $f$. The fitting is done by a genetic algorithm \texttt{PIKAIA}
\citep{charbonneau_genetic_1995}. We ran 20,000 optimization runs with different
random number generator seeds and each run would return a slightly different
optimal solution. We then fitted Gaussians to the distributions of output
parameter values and taking the mean of these Gaussian fits as the most probable
parameter value. These values are shown in Table~\ref{leastsquarefit} along with the values of the least squares fit. The best fit provided by \texttt{PIKAIA} is perfectly consistent with the values found by least square fitting, but we also get a value for the additional mass ratio parameter, which we cannot obtain directly from the data. Surprisingly, the most probable value is 1/2 which means that the two stars carry the exact same mass. The total stellar mass that we obtain here, purely from the determination of the apparent orbital period, is very consistent with the mass of 1.9 M$_\odot$ obtained by \citet{brinch_interplay_2013} based entirely on radiative transfer modeling of the velocity field in the circumbinary gas.

\begin{figure}
\includegraphics[width=8.5cm]{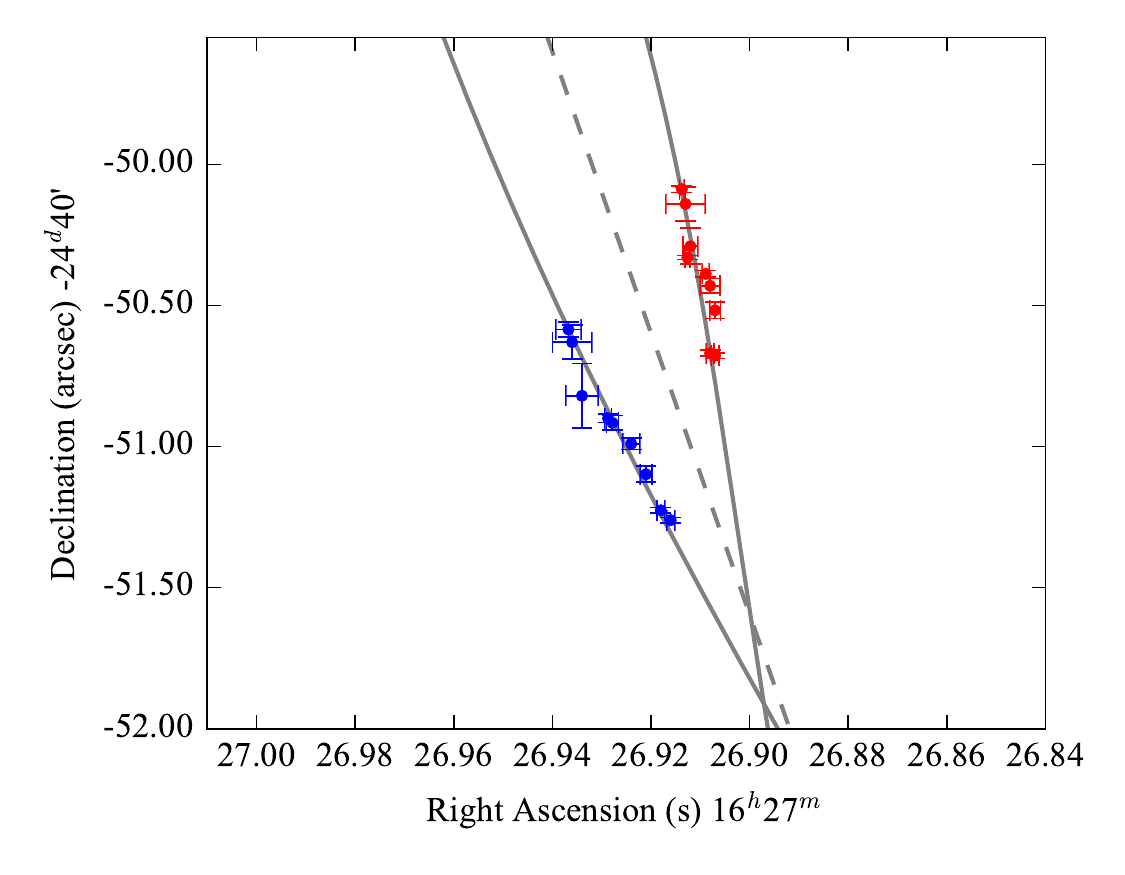}
\caption{Measured positions of VLA 1 (red) and VLA 2 (blue). The grey lines show
the best fit trajectories while the broken line shows the trajectory of the
center of mass.}\label{trajec}
\end{figure}

The trajectories of the two protostellar components and their measured positions
on the sky are shown in Fig.~\ref{trajec}. The dashed line shows the trajectory
of the center of mass.

\section{Discussion and conclusions}\label{disc}
The most curious misalignment in the IRS~43 system is that of the circumbinary disk and the orbit of the binary itself. The orbital plane of the binary is strongly constrained by the constant separation of VLA 1 and VLA 2 in time to be close to the plane of the sky ($i<30^\circ$), whereas the circumbinary disk is seen nearly edge-on. Not only does the aspect ratio of the circumbinary emission distribution prevent a line of sight inclination less than $\sim$ 60$^\circ$, but the PV-diagram (Fig.~\ref{pv_hcn}, left) is also very well described by a Keplerian profile using a mass of 1.8 M$_\odot$ and an inclination of 70$^\circ$. This inclination is consistent with the inclination found by \citet{brinch_interplay_2013}. If we use the slightly higher total mass of 2.01 M$_\odot$ found by the \texttt{PIKAIA} fit, we need an inclination of 60$^\circ$ to fit the PV-diagram. The red side of the HCN line has a strong non-Keplerian peak at a radius of 300 AU. This peak masks the Keplerian signature on the red side at this radius, but the emission actually follows the Kepler profile, although at a very low signal to noise, out to about 600 AU or slightly more. It is not clear what causes the bright spot on the red side at 300 AU.

Similarly, we can fit the PV-diagram of the circumstellar disk around VLA 1 using a mass of 0.9 M$_\odot$ and an inclination of 33$^\circ$ (30$^\circ$, if we use 1.005 M$_\odot$). We cannot determine the line of sight inclination of the circumstellar around VLA 2 due to low singal to noise, but this disk has a velocity gradient which is offset by almost 90$^\circ$ with respect to the other two disks. The orientation of the disk around VLA 2 is further supported by the linear structure in HCO$^+$, most likely an outflow, emerging from VLA 2 toward North-West and almost perpendicular to the velocity gradient.

The warp in the disk, most apparent in the blue-shifted HCN emission, is expected from simulations \citep{larwood_hydrodynamical_1997} and it is also to be expected that a cavity in the circumbinary disk be carved by the binary \citep{artymowicz_dynamics_1994}. Whether such a cavity can be seen in our data is not clear, but there is a hint of a drop in HCN intensity in the central parts of the circumbinary disk. Higher resolution observations by ALMA could reveal this.

The significant misalignment of this system suggests a formation scenario either driven by turbulent fragmentation~\citep{padoan_stellar_2002} or that the system has recently been perturbed, e.g., by ejection of a third stellar component. Another possibility is that the misalignment between the central binary and the circumbinary disk may simply be a signature of formation of the binary in a cloud, with infall of material with misaligned angular momentum, occurring late in the formation of the circumbinary disk. This would imply that the material from which the system formed was not in solid body rotation as is often considered the initial condition for star formation, but rather that the formation was triggered in a cloud with significant turbulence or substructure.

The question is whether IRS~43 will eventually align and form a planetary system which is similar to all the well aligned systems found by Kepler. \citet{bate_alignment_2010} showed that, in particular for light disks, a misalignment of the rotation axis of a protostar and its disk may survive for the entire life time of the disk. In the case of IRS~43, alignment of disks with binaries, tend to happen on the viscous time scale \citep{papaloizou}. For the two circumstellar disks, at the radius of 25 AU, we estimate the viscous time scale to be between $8\times 10^5$ and $8\times 10^6$ years for values of the $\alpha$-viscosity parameter \citep{shakura} between $10^{-2}$ and $10^{-3}$. This system could take a very long time to align ($\sim 10^6$ years), possible beyond the disk life times and certainly longer the the present age of IRS~43 \citep[$\sim 10^5$ years,][]{brinch_interplay_2013}.

\acknowledgements This paper makes use of the following ALMA data:
ADS/JAO.ALMA\#2013.1.00233.S and ADS/JAO.ALMA\#2013.1.00955.S. ALMA is a
partnership of ESO (representing its member states), NSF (USA) and NINS
(Japan), together with NRC (Canada), NSC and ASIAA (Taiwan), and KASI
(Republic of Korea), in cooperation with the Republic of Chile. The
Joint ALMA Observatory is operated by ESO, AUI/NRAO and NAOJ.

\end{document}